\documentstyle{article}

\begin{document}
\title{
Higgs field as the gauge field corresponding to parity in the
usual space-time }  
\author{Recai Erdem}
\maketitle
\small
\begin{center}
Department of Physics\\
Izmir Institute of Technology\\
G.O.P. Bulvari No.16\\
Izmir, Turkey
\end{center}
\normalsize
\date{\today} 
\begin{abstract}
We find that the local character of field theory requires the parity 
degree of freedom of the fields to be considered as an additional
discrete fifth dimension which is an artifact emerging due to the 
local description of space-time. Higgs field can be interpreted as the 
gauge field
corresponding to this discrete dimension. Hence the noncommutative geometric
derivation of the standard model follows as the manifestation of the 
local description of the usual space-time.  
\end{abstract}
\section{Introduction}
Higgs field can be considered as an additional gauge field in a generalized
covariant derivative. This identification is usually realized by identifying
it by a connection either corresponding to an extra continuous [1] or 
discrete [2] dimension or corresponding to extension of the usual space-time
by addition of a finite number of points either in a non-commutative 
differential geometry context [3] or in the usual differential geometry 
supplemented by some physical considerations [4]. Yet another way to 
obtain the
generalized covariant derivative including the Higgs field is to make use
of an axiomatic approach [5]. There are also some variants and improvements
of these formulations [6]. The 
unpleasant feature of the approaches
using an extended space-time is that there is no sound  verification for 
such an enlargement of the space-time. Although the axiomatic 
approach is attractive in deriving the nice aspects of the scheme in
the usual space-time it lacks
a deeper understanding both mathematically and physically. 

In this study
we shall see that the noncommutative differential geometry setting 
directly follows from the locality of field theory in space-time. We find 
that the discrete space-time transformations can be incorporated into the 
formulation of field theory as a fifth discrete dimension corresponding
to parity. We observe that in fact this discrete fifth dimension is hidden
in the Dirac equation. After identification of the derivative and the
gamma matrix corresponding to this dimension and after incorporating
the usual gauge group structure we derive the noncommutative geometrical
setting. In summary we show that the noncommutative
geometrical framework is hidden in the usual space-time structure and it
emerges as an artifact of field theory due to its insufficiency to include
discrete space-time degrees of freedom through the usual continuous 
space-time coordinates
.
\section{Local description of space-time and the need for the fifth 
discrete dimension}

Assume that we define a field, $\phi$ over a differentiable manifold $M$.
Provided one knows $\phi (\zeta)$ at every point $\zeta\in M$ one
knows every information about it. In other words in an action at a distance
formulation one does not need to know $\partial_{\zeta}\phi$ ( or more 
generally $D_{\zeta}\phi$) etc. to determine $\phi$ at every point. On the
other hand if one uses a local formulation (i.e a field formulation) then
one needs to know the variation of $\phi$ with $\zeta$ as well, that is,
\begin{equation}
\phi (\zeta )=\phi (\zeta_{0})+(\zeta - \zeta_{0})
(\partial_{\zeta}\phi (\zeta ))_{\zeta = \zeta_{0}}+.......
\label{a1}
\end{equation}
One can write $M$ as a direct product of two manifolds, one expressing the
internal structure, $I$ and the other the space-time structure, $S$;
$M=I\otimes S$. If we express $S$ (i.e the space-time) in a local (field) 
formulation then we can expand $\phi$ in a Taylor expansion as
\begin{eqnarray}
\phi (\rho,\eta )=&&\phi (\rho,\eta_{0})+(\eta - \eta_{0})
(\partial_{\eta}\phi (\rho,\eta ))_{\eta = \eta_{0}}+....... \nonumber \\
=&&\phi (\rho,\eta_{0})+(x- x_{0})
(D_{x}\phi (\rho,x))_{x = x_{0}}+....... \nonumber \\
&&\rho\in I~,~~~\eta\;,\eta_{0}\in S~,~~~x,x_{0}\in X
\label{a2}
\end{eqnarray}
where $X$ is the tangent space at $\eta$ and $D$ is the covariant derivative
corresponding to the use of $X$ as the local coordinate frame.

Physical observables are the quantities which do not change under 
reparametrization of $\zeta$ (i.e under symmetry transformations). In other
words they are invariants under symmetry transformations. They can be 
constructed as
\begin{equation}
I={\cal T}[L_{\phi}(\phi, D\phi)L_{\zeta}(\zeta , \delta \eta)] 
\label{a3}
\end{equation}
where $L_{\zeta}$ represents the physical observables related to the symmetry
transformations independent of $\phi$ and ${\cal T}$ denotes a generalized
trace including integration. In the case of the usual 4-dimensional
space-time coordinates
\begin{equation}
I=\int {\cal L}(\phi, D\phi)\sqrt{det(g_{\mu\nu})}d^{4}x \label{a4}
\end{equation}
where $g_{\mu\nu}$ is the metric tensor.
  
Does Eq.(\ref{a4}) really define the usual (Minkowski) space-time wholly?
It includes $\phi$ and $\partial_{\mu}\phi$ $\mu =0,1,2,3$ so it describes 
the behaviour of $\phi$ under translations, rotations, Lorentz boosts etc.
i.e the usual continuous space-time transformations hence the corresponding
physical observables. In other words the generators of the continuous 
space-time transformations can be expressed in terms of the differetial 
variation of the usual 4-dimensional coordinates, $x_\mu$. What about the 
discrete space-time transformations,
parity and time reversal? If we do not include the spinor representaions 
or if we do not introduce the notion of intrinsic parity then the 
degrees of freedom corresponding to parity and 
time reversal transformations are automatically included in the 
Lagrangian because, in this case, once we impose Lorentz invariance the 
Lagrangian becomes parity and time reversal invariant.  
Otherwise (i.e. if we admit spinor representations of the Lorentz group 
with an internal group structure or admit 
intrinsic parity) they are not included in the above formulation
unless they are included in an ad hoc way because one can not generate them 
from the continuous transformations. Then there are three alternatives
either one uses an action at a distance formulation or one includes their
effect in an ad hoc way ( i.e without a differential geometric formulation)
or one should extend the differential geometric formulation so that these
discrete degrees of freedom are included as the local rate of change of 
$\phi$ under these transformations. In other words one must introduce an
additional dimension corresponding to these degrees of freedom which
disappear when one passes to an action at a distance formulation.

\section{Parity as the fifth local dimension}

We can summarize the previous section as follows: If one uses an action at
a distance formulation it is enough to know $\phi(\zeta)$ $\zeta\in M$
for all $\zeta$ in order to have all information about the state 
described by $\phi$ but when one uses field formalism one must know 
$\phi(\zeta)$ and $\phi(\zeta+\delta\zeta)$ at some neigborhood of
$\zeta$ or equivalently $\phi(\zeta)$ and $D\phi(\zeta)$. In the case
of parity this is equivalent to saying that, in field formalism if we 
include the spinor reresentations of lorentz group with an internal group 
structure or if we admit intrinsic parity, one must know $\phi(\vec{x},t)$ 
and $\phi(-\vec{x},t)$ or 
equivalently $\phi(\vec{x},t)$ and $D_{p}\phi(\vec{x},t)$ at the point
$(\vec{x},t)$. One can define the sets
\begin{equation}
X_{L}=\{\vec{x},t\;|\;\vec{x},t\in 
S\}~~~X_{R}=\{-\vec{x},t\;|\;\vec{x},t\in S\} \label{g1}
\end{equation}
or in a form that reflects local
describtion of space-time
\begin{equation}
X_{L}=\{d\vec{x}, dt\; |\; \mbox{for all}~~ d\vec{x},dt\in O\}~~~
X_{R}=\{-d\vec{x},dt\;|\; \mbox{for all}~~ d\vec{x},dt\in O\}
\label{g2}
\end{equation}
where $O$ is a neigborhood of $\vec{x},t$. Parity can be considered
to be an extra direction connecting $X_{L}$ and $X_{R}$ in addition
to the usual directions $x_{\mu}$ $\mu = 0,1,2,3$. This is due to the
fact that the information about $\phi(\vec{x},t)$ and 
$\phi(\vec{x}+d\vec{x}, t)$ does not tell anything about the behaviour of
the field under parity because one can not generate parity through 
continuous transformations. When we only consider the behaviour
of the field under parity transformation one can consider $X_{L}$ and
$X_{R}$ as two discrete points in this additional direction. Below
we shall formulate these observations in a more systematical way. 

We consider the parity degree of freedom of the fields to correspond to
a fifth discrete dimension consisting of two points
\begin{eqnarray}
p_{1}=\left(\begin{array}{c}
0\\
1\end{array}\right)~~~~
p_{2}=\left(\begin{array}{c}
1\\
0\end{array}\right)
 \label{b1}
\end{eqnarray}
which are connected by parity transformation
\begin{equation}
P: p_{1(2)}=p_{2(1)} \label{b2} ~.
\end{equation}
One notices that in fact $p_{1}$ and $p_{2}$ belong to different dimensions
otherwise when a left-handed particle is created its partner right-handed
would be destroyed. ( But we shall see later that effectively there is 
only one dimension)

We shall put a two-component left-handed spinor at point  
$p_{2}$ and put a right-handed one at point $p_{1}$ , that is,
\begin{eqnarray}
&&\psi (p)=\left(\begin{array}{c}
\psi_{1}(p_{2})\\
\psi_{2}(p_{1})\end{array}\right)=
\left(\begin{array}{c}
\psi_{L}\\
\psi_{R}\end{array}\right) \nonumber \\
~~and &&\left(\begin{array}{c}
\psi_{2}(p_{1})\\
\psi_{1}(p_{2}\end{array}\right)=0~~,~~\left(\begin{array}{c}
\psi_{1}(p_{1})\\
\psi_{2}(p_{2})\end{array}\right)=0
\label{b3}
\end{eqnarray}
Under parity transformation
\begin{eqnarray}
P:\psi (p)\rightarrow \psi^{\prime}(p^{\prime})=\left(\begin{array}{c}
\psi_{2}(p_{2})\\
\psi_{1}(p_{1})\end{array}\right)=\left(\begin{array}{c}
\psi_{R}\\
\psi_{L}\end{array}\right) \label{b4}
\end{eqnarray}
while from Eq.(\ref{b3}) it follows that
\begin{eqnarray}
\psi^{\prime}(p)=\left(\begin{array}{c}
\psi_{2}(p_{1})\\
\psi_{1}(p_{2})\end{array}\right)=0 \label{b5}
\end{eqnarray}
So 
\begin{equation}
\psi^{\prime}(p^{\prime})=
\psi^{\prime}(p)+\delta_{p}\psi (p)= \delta_{p}\psi (p) \label{b6}
\end{equation}
In other words
\begin{equation}
\delta_{p}\psi=\psi^{\prime}(p^{\prime})=\gamma^{0}\psi (p^{\prime}) 
\label{b70} \end{equation}
where it is understood that $\psi_{R}$ in Eq.(\ref{b3}) acts as a
left handed spinor while $\psi_{R}$ acts as a right-handed one under
SL(2,C). Now we determine the differential length in the fifth 
dimension, $dx_{4}$. 
\begin{eqnarray}
dx_{4}&&=dx_{p}=\frac{1}{im}(p_{2}-p_{1})=\frac{1}{im}(
\left(\begin{array}{c}
1\\0 \end{array}\right)
-\left(\begin{array}{c}
0\\1 \end{array}\right)) \nonumber \\
&&=\frac{1}{im}
\left(\begin{array}{c}
1\\-1 \end{array}\right)=\frac{1}{im}\gamma_{5}\gamma^{0}
\left(\begin{array}{c}
1\\1 \end{array}\right) \label{b8}
\end{eqnarray}
Although the use of $\gamma^{0}$ may seem redundant in fact it is essential
in order to match the differential distances in
$x_5$ with the correct elements of $\psi$
because 1 in $(1,-1)^T$ corresponds to the change in the coordinates of
$\psi_{L}$ which moved to lover position in $\delta\psi$ and vice versa for
-1 and $\psi_{R}$.
Therefore the derivative operator corresponding to the fifth dimension is
found to be
\begin{equation}
\partial_{4}=\frac{d_{4}}{dx_{4}}=im\gamma^{0}(\gamma^{0}\gamma_{5})
=im\gamma_{5} \label{b9}
\end{equation}
One can identify $i\gamma_{5}=\gamma^{4}$ as the gamma matrix 
corresponding to the fifth dimension because
\begin{eqnarray}
\{\gamma^{\mu},\gamma^{\nu}\}=2\,g^{\mu\nu}&&~~\mu ,\nu =0,1,2,3,4 
\nonumber \\
&&(g^{\mu\nu})=diag(1,-1,-1,-1,-1) \label{b10}
\end{eqnarray}
One recognizes that in fact the fifth dimension corresponding to 
parity transformations is hidden in Dirac equation
\begin{equation}
\not{D}\psi -m\psi=\not{D}\psi -\gamma^{4}\partial_{4}\psi=0 \label{b11}
\end{equation}
Although the presence of $\gamma^{4}\partial_{4}$ has no effect
here we shall see in the next section that this is not the case when
$\psi$ has a local internal group structure.

One can also derive the Dirac action corresponding to Eq.(\ref{b11})
through this formalism
\begin{equation}
S_{D}=im\int \psi^{\dagger}dx_{4}\gamma^{\mu}\partial_{\mu}\psi =
\int \bar{\psi}(\not{partial}-m)\psi\,d^4x ~~~~\mu = 0,1,2,3,4 \label{f1}
\end{equation}
where we have used the identity
\begin{eqnarray}
&&\int_{A}^{B} 
\psi^{\dagger}\,dx_{4}\gamma^{\mu}\partial_{\mu}
\psi =
\int_{A}^{B}\gamma^0\gamma_{5}\gamma^{\mu}\partial_{\mu}\psi\,|dx_{4}|
\nonumber \\
&&=
\int_{a}^{b}\psi_{L}^{\dagger}\gamma^{\mu}\partial_{\mu}\psi_{L}|dx_4|-
\int_{b}^{a}\psi_{R}^{\dagger}\gamma^{\mu}\partial_{\mu}\psi_{R}|dx_4|=
\int{a}^{b}\bar{\psi}\gamma^{\mu}\partial_{\mu}\psi\,|dx_4|
\label{f3}
\end{eqnarray}
 
This construction can be understood better through a mathematical
analysis. The above gamma matrices in five dimensions correspond to 
spinorial representations of $SO(5)$.
Because the spinorial representations of both $SO(2k)$ and $SO(2k+1)$
are $2^k$ dimensional the spinorial representations of both $SO(5)$ and
$SO(4)$ are 4-dimensional. Hence the spinorial representations
of the vectors in 5 dimensional pseudo-Euclidean space can be determined
by adding another independent $2^2$ dimensional matrix (i.e $\gamma^4$)
to $\gamma^{\mu}$'s of 4-dimensional Minkowski space [7]. In fact this
is realized another form as well: The gamma matrices of 4-dimensional
Eucleadian space and the generators of $SO(4)$ together form the
generators of SO(5) [8]. 

Another discrete space-time transformation is time reversal [9]. Fortunately
time reversal does not enter as an additional discrete dimension once
parity is taken to correspond to the fifth dimension because one can
generate a time reversal through a rotation in plane plus an analytic
extension (!) of Lorentz boosts. For example
\begin{eqnarray}
&&P:(x_0,x_1,x_2,x_3)\rightarrow (x_0,-x_1,-x_2,-x_3)~~~
R:(x_0,-x_1,-x_2,-x_3)\rightarrow (x_0,-x_1,x_2,x_3) \nonumber \\
&&\Gamma :(x_0,-x_1,x_2,x_3)\rightarrow (-x_0,x_1,x_2,x_3) \label{b12}
\end{eqnarray}
where
\begin{eqnarray}
&&P=\mbox{diag}(1,-1,-1,-1)~~~
\Gamma =\mbox{diag}(-1,-1,1,1) \nonumber \\
&&R=\left(\begin{array}{cccc}
1&0&0&0\\
0&1&0&0\\
0&0&-1&0\\
0&0&0&-1\end{array}\right) \label{b13}
\end{eqnarray}
Here $\Gamma$ is an analytic extension of the usual Lorentz 
transformations with
\begin{equation}
\gamma =\sqrt{1-\beta^2}=-1 ~~\mbox{for}~~\beta =0 \label{b14}
\end{equation}
Although this choice is not physical it is a legitmate one because in the
differential geometrical construction of the action functional we consider
all kinds of variations of the variables. In other words this choice is 
already included in the construction of the action. Therefore although 
time reversal is an independent space-time transformation it does not 
introduce a new artificial dimension. (Remember that the fifth dimension
itself is not true dimension. It is an artifact imposed by our insistence
of a wholly differential geometric construction of action functional.) This 
fact can be
understood in another way also: Both parity and time reversal interchange
the dotted and the undotted representations of SL(2,C) so both act on the
same discrete dimension. Of course when we consider an internal structure
for the fields then time reversal and charge conjugation operations can
be considered to correspond to some symmetries whose connection can be 
identified with another set of Higgs fields. But this connection is 
different from the one corresponding to a parity transformation
because the laters correspond to the extension of the usual gauge group while
the fifth dimension can be identified as gauge field without extending the
gauge group.

\section{Inclusion of internal structure and emergence of Higgs field}
 
Above we have seen that the (formal) fifth dimension is hidden in the Dirac
equation if we only consider space-time. When we consider an internal
structure for the fields then the fifth dimension becomes more explicit.
For example in the standard model (considering only the weak sector)
\begin{eqnarray}
&&\partial_{4}=\frac{d_{4}}{dx_{4}}=m(\gamma^{0}\gamma^{4})\gamma^{0}
\rightarrow (M\otimes \gamma^{0}\gamma^{4}) (\tau_{1}\otimes 
I_{4}\otimes I_{3}\otimes \gamma^{0})\nonumber \\
&& =-M(\tau_{1}\otimes 
I_{4}\otimes I_{3})\otimes 
\gamma^{4} \nonumber \\ 
&& \gamma^4 \partial_{4}=M(\tau_{1}\otimes I_{4}\otimes 
I_{3})\otimes I_{4}  \label{c1} \end{eqnarray}
where $M=$diag$(M^{\prime},M^{\prime\dagger})$ $[M^{\prime}$=
diag$(M_{u},M_{d},0,M_{e})]$ is the fermion mass matrix.
( Note that the unusual $\gamma_{5}$ in the standard formulation of the
noncommutative geometry [10] does not appear here because we use use 
Minkowski signature instead of Euclidean one.) The fermion mass matrix, 
$M$ in Eq.(\ref{c1}) 
can be taken to be the generalization of the fermion masses in the usual
Dirac equation. It can be better understood if we consider it to correspond
to the increase of the discrete points in the fifth dimension  from 2 to 
$2\times 4\times n=8n$ while the space-time is still five 
dimensional. We can assume that each $\psi_{iaL(R)}$, 
($a=u,d,\nu ,e$ and $i=1,2,...n$ for $n$ families of fermions) has a 
differential variation
in the fifth dimension given by $m_{ia}$ when $\psi_{iaL(R)}$ is in the 
mass basis. Then the fermion mass matrices are found from
\begin{equation}
\bar{\psi}_{ia}(\partial_{4}\psi_{ja})+(\bar{\psi}_{ia}\partial_{4})
\psi_{ja}=\bar{\psi}_{iaL}(x_{ija}+x_{jia}^{*})\psi_{jaR}+h.c.=
\bar{\psi}_{aL}U_{L}^{\dagger}\,m_{a}\,U_{R}\psi_{aR}+h.c.
\end{equation}
that is $M_{ija}=x_{jia}^{*}+x_{ija}$ where a non-diagonal $M_{a}=
U_{L}^{\dagger}\,m_a\,U_R$ 
corresponds to the general case where $\psi_{ia}$'s are not in the mass
basis i.e. each $\psi_{ia}$ does not sit at one point in the fifth 
dimension but rather they are at general points in flavor space which
can be expressed as admixtures of $1/m_{ia}$'s.

\section{Conclusion}

In this study we have seen that parity can be incorporated into the 
differential geometric construction of field theory as a fifth discrete
dimension otherwise while the continuous part of space-time transformations
are included in an elegant differentail geometric formulation 
the discrete transformations can be incorporated in an ad hoc way 
into the local 
formulation of the field theory . 
We have shown that the differentiation along the fifth discrete direction is 
nontrivial
especially if we include an internal gauge structure for the fields. The
corresponding connection can be identified as the Higgs field of the
standard model if take the gauge group as 
$SU(3)_{c}\otimes SU(2)_{L}\otimes U(1)$. In other words Higgs field
can be considered as a gauge field without extending  the (Minkowski)
space-time and the ($SU(2)_{L}\otimes U(1)$) gauge group. We hope that
the studies in this direction supplemented with grand unification
schemes of multiple intermediate symmetry breaking scales will
contribute to a more unified, more predictive and renormalization group
invariant account of particle physics [5,11].


\begin{thebibliography}{99}
\bibitem{}  N.S. Manton, Nucl. Phys. {\bf B158} (1979) 141
\bibitem  B.S. Balakrishna, F. G{\"u}rsey and K.C. Wali,
 Phys. Lett. {\bf B254} (1991) 430 \\
R. Coquereaux, G. Esposito-Farese and G. Vailant, 
Nucl. Phys. {\bf B353} (1991) 689 
\bibitem{}  A. Connes, Essay on Physics and Non-commutative Geometry, 
{\it in} Interface of Mathematics and Physics, ed. D.G. Quillen, G.B. 
Segal and Tsou S.T. (Oxford Univ. Press, Oxford, 1990)
 \item {} G.K. Konisi and T. Saito, Prog. Theor. Phys. {\bf 95} 
(1996) 657; \\
A. Connes and J. Lott, Nucl.Phys. B (Proc. Suppl.) {\bf 18 B} (1990) 29; \\
T. Sch{\"u}cker and J.-M.Zylinski, J. Geom. Phys. {\bf 16} (1995) 207; \\ 
\bibitem{}  I.S. Sogami, Prog.Theor. Phys. {\bf 95} (1996) 657
\bibitem{}  A.H. Chamsedine, G. Felder and J. Fr{\"o}lich,
Phys. Lett. {\bf B296} (1992) 109; ibid, Nucl. Phys. {\bf B395} (1993) 672;\\
A. Sitarz, Phys. Lett. {\bf B395} (1993) 672; \\
K. Morita and Y. Okumura, Prog. Theor. Phys. {\bf 91} (1994) 959;\\
S. Naka and E. Umezawa, Prog. Theor. Phys. {\bf 92} (1994) 225
\bibitem {} F.D. Murnaghan, The Theory of Group Representations, 
( Dover, New York, 1963) 
\bibitem {} E.M. Corson, Introduction to Tensors, Spinors, and 
Relativistic Wave-Equations, ( Blackie{\&}Son, Glasgow, 1953)  and the 
references therein\\
H.J. Bhabha, Rev.Mod.Phys. 17 (1945) 200 and Proc.Indian Acad.Sci 21 
(1945) 241, {\it in} Homi Jehanger Bhabha, Collected Scientific Papers,
( Tata Institute Fundamental Research, India, 1985) 
\bibitem {} R.G Sachs, The Physics of Time Reversal, 
(University of Chicago Press, USA, 1987)  
\bibitem {} D. Kastler and T. Sch{\"u}ker, CNRS Preprint
CPT-94/P.3091, hep-th/9412185, 1994 
\bibitem{} E. Alvarez, J.M. Gracia-Bondia and C.P. Martin, Phys. Lett.
{\bf B306} (1993) 55; ibid, Phys.Lett. {\bf B329} (1994) 259
 
\end{thebibliography}
\end{document}